\documentstyle[11pt]{article}
\newcommand{\be}{\begin{eqnarray}}
\newcommand{\ee}{\end{eqnarray}}
\title{\begin{flushright}
{\small SUNY-NTG-94-4\\
NSF-ITP-93-150}
\end{flushright}
{\bf QUARK-INDUCED CORRELATIONS BETWEEN INSTANTONS
DRIVE THE CHIRAL PHASE TRANSITION}}
\author{{\bf E.-M.Ilgenfritz  }$^{\dagger}$ and
 {\bf E. V. Shuryak}$^{\ddagger}$ \\
  {\it Institute for Theoretical Physics,}
  {\it University of California} \\
  {\it Santa Barbara, CA 93106}}

\begin{document}

\maketitle
\centerline{\bf Abstract}

A simple model for the instanton ensemble at finite temperature T is proposed,
including ``random" and strongly
correlated ``molecular" component.
 T-dependence of fermionic zero modes
 naturally leads to chiral symmetry restoration, without
 instanton suppression. Moreover,
at $T=(1-2) T_c$ the non-perturbative effects
due to ``molecules" are  so strong, that they even
 dominate the global thermodynamics.

\noindent $^{\dagger}$ On leave from
{\it Fachbereich Physik, Humboldt Universit\"at Berlin, Germany}

\noindent $^{\ddagger}$ On leave from: {\it Physics Department,
State University of New York, Stony Brook, NY 11794}

\newpage
  Instantons are the major component of non-perturbative fields in the
QCD vacuum, and significant amount of work
 \cite{CDG,Shuryak_82,DP}
has been done
in order to build a quantitative theory describing them.
 Two major steps forward were done during the last year.
 First, the simplest ensemble (the so called
Random Instanton Liquid Model), has  reproduced many correlation
functions \cite{SV} known from phenomenology
 \cite{Shuryak_cor}
and lattice simulations \cite{Negele}.
Second, by ``cooling" of the (quenched) lattice configurations
it was found in \cite{Negele_DALLAS}  that the
 typical instanton density is about $n\approx
1.4 fm^{-4}$ and the typical size is about $\rho\approx .35 fm$, very close to
the ``instanton liquid"
parameters suggested
by one of us a decade ago \cite{Shuryak_82}.
The correlation functions and hadronic
 wave functions,
are shown to be practically unaffected by ``cooling": so,
by removing {\it perturbative gluons} and
{\it confinement}, one does not
loose mesons and baryons!

   The instanton-induced effects at finite temperature $T$
and much less studied. Our particular focus is on
 the
chiral restoration
phase transition
at $T_c$. The main idea is that it happens
due to growing correlations between instantons
 and anti-instantons,
described by the disappearance of ``single" instantons and
 growth of $\bar I I$ ``molecules".

   Our first study along this line was performed  few years ago
 \cite{IS}. However,
  the present paper is significantly different, because it is based on
 completely new mechanism. Instead of
Debye-type screening \cite{Shuryak_conf}, implemented by
{\it thermal suppression factor} $f(T)=exp(-\rho^2 T^2 const)$
\cite{PY} \footnote{This suppression is expected to work at {\it high} $T>>T_c$
only, and available
lattice data on screening mass suggest
that it may probably happens at only at temperatures $T>300-400 MeV$.
Direct
lattice measurements of the instanton density \cite{DiGiacomo} have
indeed seen
no significant $T$ dependence till such high temperatures.
}
, the phase transition occurs due to the T-dependence of the
 {\it quark-induced}
$\bar I I$ interaction.

The ensemble of interacting instantons can be studied
with the partition function
\be
Z= \int \sum_{N_+ N_-} {1 \over N_+ ! N_- !}
   \prod_i^{N_+ + N_-} d\Omega_i d(\rho_i) \rho_i^{N_f}
exp(-S_{int})\Pi_i^{N_f} det(i\hat D+im_f) \ \ \,
\ee
where $d\Omega_i$ is the measure in space of collective coordinates,
(12 per instanton in $SU(3)$), $d(\rho)$ the instanton amplitude, and $S_{int}$
the gluonic interaction. The last factor, appearing after integration
over fermions, is
the one we are focussing on. Assuming $N_+=N_-=N$,
one writes it in terms of a $N\times N$ 'hopping' matrix, with
$ T_{IA}=\int d^4x \langle \psi_{A0}(x-z_A)\vert i\hat D_x \vert
\psi_{I0}(x-z_I) \rangle$
from some instanton $I$ to some
anti-instanton $A$
($z_A,z_I$ are the centers of $A,I$, $\psi_0$ the zero-modes),
as
$ det(T T^+ + m_f^2) $. The
statistical system described by this partition function
is quite complicated, and direct
simulations were done so far only for $T=0$.

   The instanton solutions and their zero modes are known analytically for
non-zero $T$, and detailed studies of 'hopping' matrix elements
were done in \cite{SV_T}. They have the structure
 $ T_{IA}=u_4 f_1+ ({\bf u  r}/ r) f_2 $,
where the $2\times 2$ matrix
 $u_\mu\tau_\mu^+$ describes the relative orientation of
 $I$ and $A$.
Rather complicated
formulae for $f_1,f_2$  were derived in \cite{SV_T}.

We consider the instanton ensemble as
a superposition of the uncorrelated (or ``random'") component and a
highly correlated (or ``molecular") component, with the
(4-d) densities $n_a(T),n_m(T)$. In the former component
instantons are assumed to have {\it random} relative orientation $u_\mu$, and,
 as in \cite{IS}, it is treated in a
mean field approximation. The fermionic
determinant is obtained from  $I(T)\sim\langle T_{IA} T^+_{AI} \rangle$,
{\it i.e.} summed over anti-instantons $A$ with a density $n_a/2$,
with random {\it positions} and {\it orientations}.

 For the ``molecular" component
we assume the opposite, namely the most favorable
relative color orientation $u_\mu \sim (z^\mu_I-z^\mu_A)$,
maximizing the
hopping matrix element $|T_{IA}|$ (as well as  $exp(-S_{int})$).
Therefore, we have to calculate
another function $\tilde I(N_f,T)\sim\langle (T_{IA} T^*_{AI})^{N_f} \rangle$
where now averaging
  means integration only over the {\it relative coordinates}
inside a $\bar I I$ pair.

   Temperature dependence of these average matrix elements was obtained
 by numerical integration of the formulae from \cite{SV_T}, and the results
are shown in Fig. 1. One has to exclude too close
 $\bar I I$ pairs\footnote{The general reason for such 'repulsive core'
is generally related to the fact, that too close pairs do not in fact
correspond to strong fields and are not objects of the semiclassical
theory.}
, and we show results with the ``core radii" $R_c=1\rho$ (dashed)
and  $R_c=2\rho$
 (solid).  Although
the overlap integrals significantly depend on $R_c$,  the
resulting uncertainty of
the thermodynamical quantities (see below) is in fact not so dramatic.
  Note remarkably
different $T$-dependence
of these two quantities. While the $I(T)$ (indicated as the $N_f=1$ curves
in Fig.1) {\it decreases} with $T$, $\tilde I(N_f,T)$
 {\it grows}, the stronger the
larger  $N_f$. As a result, the molecular component builds up with $T$, while
the random one decreases and eventually disappears, restoring
the chiral symmetry.

   In order to simplify gluonic interaction, and still
describe the system self-consistently, we adopt a simple ``average repulsion"
$ \langle S_{int} \rangle =\kappa \rho_I^2 \rho_A^2 $
with one dimensionless parameter $\kappa$.
The same parametrization is used for the
description of interaction between all instantons,
belonging to the
random component or to molecules.

  Let us now evaluate the statistical sum of the system in terms of
the densities
$n_a,n_m$, starting with the differential activities for molecular
\be
dz_m=C^2 d\rho_1 d\rho_2 d^4R {d\Omega_{SU(3)} \over\Omega_{SU(3)}}
(\rho_1\rho_2)^{b-5} exp[-\kappa(\rho^2_1+\rho^2_2)(\bar{\rho^2_a}n_a
+2\bar{\rho^2_m}n_m)] (T_{IA}T_{AI}^*)^{N_f}
\ee
and random components
\be
dz_a = 2 C d\rho \rho^{b-5} exp[-\kappa \rho^2 (\bar{\rho^2_a}n_a
+2\bar{\rho^2_m}n_m)] <TT^+>^{N_f}
\ee
where $b={11\over3}N_c-{2\over 3}N_f$ is the coefficient of the Gell-Mann-Low
function and\\
$<TT^+>=\rho^{3/2} [{1\over 2} I(T) \int dn_a(\rho) \rho]^{1/2}$.
Besides the densities, an important ingredient of the interaction are the
 root mean square radii $\bar\rho_a,\bar\rho_m$, which can be found from
eqs. (3) and (4)
to be related to each other through
\be {\bar\rho^2_m \over \bar\rho^2_a} = {\alpha \over \beta}; \, \, \,
\alpha=b/2-1 ; \,\,\,\,\, \beta=b/2+3N_f/4-2
\ee
Another relation between them
connects the interaction parameter $\kappa$ to the
diluteness of the ensemble
\be {1 \over \kappa}= {2 \bar\rho^4_a n_a \over \beta}+
{4 \bar\rho^4_m n_m \over \alpha} \ \ \ . \ee
So, one can eliminate the mean square radii and get the
activities
\be
z_m={A \over [n_a+(2\alpha/\beta) n_m]^\alpha};
\,\,\,\, A={\tilde I(N_f,T) C^2 \Gamma^2(\alpha) \over (4\kappa\beta)^\alpha}
\ee
\be
z_a={B n_a^{N_f/2} \over [n_a+(2\alpha/\beta) n_m]^{\beta/2+N_f/8}};
\,\,\,\, B={ C \Gamma(\beta) \over (2\kappa)^\beta} ({I(T)\over 2})^{N_f/2}
({\beta\over \kappa})^{N_f/8-\beta/2}  \ \ \ .
\ee
As usual, the grand potential
\be
\Omega=-p=-(logZ)/V_4= {N_a\over V_4} log({e z_a V_4 \over N_a}) + {N_m
\over V_4} log({e z_m V_4 \over N_m} )
\ee
(where $V_4$ is the 4-dimensional volume)  should then be
minimized with respect to the particle numbers $N_a=V_4 n_a$ and $N_m=V_4 n_m$.
 The resulting grand potential provides the {\it instanton contribution}
to the pressure $p$ and to the energy density $\epsilon=-p+T{\partial p \over
\partial T}$.

 The general case leads to rather  cumbersome resulting equations, but
for $T>T_c$ one has $N_a=0$ and the following simple result
$p={b\over 2} n_m$ \footnote{ It is interesting, that connecting $n_m$
with the gluon condensate
${\alpha_s\over \pi}\langle G^a_{\mu\nu} G^a_{\mu\nu} \rangle=16 n_m$
one reproduces the famous
``trace anomaly" expression. However, that the
contribution of molecules to the energy density is not just the same
expression with the opposite sign, because the fermionic determinant has an
explicit $T$-dependence.
}.

%


   The statistical sum under consideration describes two phases,
with and without chiral symmetry.
However, the realistic description should
include  the contributions to thermodynamics {\it unrelated with instantons} as
well:  it is the {\it total} pressure, which should be continuous
through the transition. We use the simplest possible model here,
including the non-interacting
 massless pions in the broken phase (which for $N_f=2,3$ are actually
irrelevant) and the ideal quark-gluon plasma in the symmetric phase.

  A typical set of results is shown in Fig. 2
for $N_f=2$ and the cores ${R_c\over \rho} = 1$ and $2 $, to see
 uncertainties involved. The constants $A$ and $B$
 entering
the activities
could be determined from first principles, provided
we know the accurate value of $\Lambda_{QCD}$ and instanton interactions.
We select $B$ in order to get the instanton density at $T=0$ equal to that
found on the lattice, $1.4 fm^{-4}$, and $A$ in order to get $T_c \simeq
150 MeV $.

   The upper panel of Fig. 2 tells us, that
although the ``random component" (solid line)
dominates the broken phase,
the number of molecules jumps up at the transition, and the
total instanton
density  above $T_c$,  2$n_m$, turns out to be
comparable to that at $T=0$.
 We have not plotted the behaviour of the
 {\it quark condensate}, which
scales as $<\bar \psi \psi> \sim n_a(T)^{1/2}$.
Inside uncertainties of the model it is  essentially flat, till
nearly $T_c$, which qualitatively agree with lattice data.

  The importance of the molecular component is better demonstrated by
 the $T$-dependence of the pressure \footnote{
Note that both pressure and energy density are counted not from physical, but
from perturbative vacuum.}
(the middle panel of Fig. 2):
in fact, for $T=(1...2)T_c$ the contribution of
``molecules" (dash-dotted line)
is crucial, without it one would not be
able to sustain the pressure because that from quarks and gluons
 (the dotted line)
 is not large enough !

 The lower panel shows the large
jump in energy density at the transition. Although it is mostly
due to the ``liberation" of quarks and gluons,  a finite part of it
 is also generated
by  the ``molecules"
(dash-dotted line). Such behaviour was
in fact observed in lattice calculations
with dynamical fermions: unlike those for the pure glue case,
they show $\epsilon(T)$ which
overshoots the Stefan-Boltzmann value above $T_c$.

    Finally, let us discuss
  what happens at larger $N_f$. First of
 all, from (6) it is clear that a positive radius of  molecules
can only be
obtained for $\alpha>0,b>2$, or $N_f<13.5$
 \cite{largeNf}.
Furthermore, for fixed interaction (namely
for core parameter $R_c$ and $\kappa$ being independent on $N_f$)
  the broken phase  exists in a shrinking region of temperatures,
and
for $N_f \geq N_f^{upper}$ the model predicts the {\it unbroken} chiral
symmetry even in the ground state of the theory. The vacuum structure for
large $N_f$ was studied on the lattice, and
(although the question is by no means
settled) indications were reported \cite{JAPNf} that
$N_f^{upper}=7$. This number is similar to what one gets in our model.


{\bf Acknowledgements}
  Discussions of relevant  questions took place
in ITP, Santa Barbara,  during the
 'Finite temperature QCD' program. The hospitality in ITP and
the financial support by NSF under Grant No. PHY 89-04035
are greatly acknowledged. E.S. thanks M.Nowak, T.Schaefer and
J.Verbaarschot for helpful discussions.

\newpage
\centerline{\bf Figure Captions}

Fig. 1.  The ratio $\tilde I(N_f,T)/\tilde I(N_f,0)$
of the
average fermionic determinant for molecules at temperature T,
normalized to its value at T=0. Numbers on the figure are $N_f$, the number
of (massless) quark flavors. The curves marked by 1 correspond to
$I(T)/ I(0)$, the integral for ``random" component. Solid curves are for
core radius $R_c=2\rho$, and the dashed ones are for $R_c=\rho$.

Fig. 2.  Chiral restoration phase transitions for 2 massless flavors and two
different core parameters $R_c$.
The upper panel shows the $T$ dependence of the densities
$n_a$ (solid) and
$n_m$ (dotted).
In the middle panel the $T$ dependence of the total pressure $p$
(solid) is shown,
including the contributions of the pion gas/quark-gluon plasma
(dotted) and of instantons (dash-dotted).
The energy density is presented in the lower panel (solid) which is
modified by the instanton contribution (dash-dotted).
\end{document}